\documentclass{optica-article}

\journal{opticajournal} 

\articletype{Research Article}

\usepackage{physics}
\usepackage{siunitx}
\usepackage{textcomp}
\usepackage{lineno}

\begin{document}

\title{High-speed full-color computer-generated holography using a digital micromirror device and fiber-coupled RGB laser diode}

\author{Shuhei Yoshida}

\address{Department of Electrical, Electronic and Communication Engineering, Kindai University, 3-4-1 Kowakae, Higashiosaka City, Osaka 577-8502, Japan}

\email{yshuhei@ele.kindai.ac.jp} 


\begin{abstract*}
Computer-generated holography (CGH) can be used to display three-dimensional (3D) images and has a special feature that no other technology possesses: it can reconstruct arbitrary object wavefronts. In this study, we investigated a high-speed full-color reconstruction method for improving the realism of 3D images produced using CGH. The proposed method uses a digital micromirror device (DMD) with a high-speed switching capability as the hologram display device. It produces 3D video by time-division multiplexing using an optical system incorporating fiber-coupled laser diodes (LDs) operating in red, green, and blue wavelengths. The wavelength dispersion of the DMD is compensated for by superimposing plane waves on the hologram. Fourier transform optics are used to separate the object, conjugate, and zeroth-order light, thus eliminating the need for an extensive 4f system. The resources used in this research, such as the programs used for the hologram generation and the schematics of the LD driver, are available on GitHub.

\end{abstract*}

\section{Introduction}
Computer-generated holography (CGH) synthesizes holographic fringes on a computer and projects three-dimensional (3D) images in real space\cite{Lesem1968,Wyrowski1987,Slinger2005,Onural2011,Geng2013}. CGH can be considered the ultimate 3D display technology because it can produce arbitrary complex wavefronts. Full-color moving image reconstruction is used to enhance the realistic sensations produced by CGH. In addition to methods that use multiple spatial light modulators (SLMs)\cite{Sato1994,Yaras2009,Nakayama2010,Senoh2011,Sasaki2014,Zeng2017,Wang2019a}, full-color reconstruction techniques include spatial segmentation\cite{Ito2004,Makowski2010,Makowski2012,Sando2018} and time-division multiplexing using a single SLM. In time-division multiplexing\cite{Shimobaba2003,Oikawa2011,Araki2015,Wang2016,Asundi2017}, the SLM display switches quickly between holograms corresponding to red, green, and blue (RGB) wavelengths, thus producing full-color 3D images. Time-division multiplexing is also characterized by a more straightforward optical configuration than the methods that use multiple SLMs, and the imaging area is larger than for spatial segmentation. However, time-division multiplexing gives an effective frame rate of only 1/3 that of the other methods due to the switching between holograms, which may be insufficient for full-color 3D video reconstruction.

Given the above, CGH reconstruction methods using a digital micromirror device (DMD) as an SLM have been investigated. The switching speed of DMDs is about 100 times that of the widely used liquid crystal SLMs (LC SLMs), which enables full-color 3D video reconstruction at a sufficient frame rate even when using the time-division method. Details of research on full-color CGH reconstruction using DMDs can be found in Ref. \cite{Takaki2015,Matsumoto2017,Sando2018,Li2022}. In the horizontal scanning method\cite{Takaki2015,Matsumoto2017}, elemental holograms with an optically reduced pixel pitch are scanned by a galvanometer mirror and displayed at high speed to achieve a wide viewing angle in the horizontal direction. The frame rate is limited by the scanning speed of the galvanometer mirror, which is about 30 frames per second (fps). A 4f system and a single side-band filter\cite{} eliminate unwanted light, and the wavelength dispersion of the diffraction angle is compensated for by adjusting the incident angle using a fiber array. A method based on double-step Fresnel diffraction (DSFB) has also been proposed for producing full-color Fresnel holograms\cite{Li2022}. In this method, the diffraction calculation is divided into two stages: forward propagation to the filter plane and backward propagation to the hologram plane. DSFB can eliminate unwanted light without the need for a 4f system. The diffraction angles are matched by superimposing grating phases on the hologram to compensate for wavelength dispersion.

As a development of these existing methods, in this paper, a full-color 3D video reconstruction technique that uses a simple configuration based on a DMD and fiber-coupled RGB laser diodes (LDs) is proposed. This system produces full-color 3D video by switching between multiple sets of binary amplitude holograms corresponding to three wavelengths that are generated by the polygon method\cite{Matsushima2005,Matsushima2009,Pan2013,Zhang2022} at high speed. The DMD switching speed determines the video frame rate; in principle, the system can produce 3D video at speeds exceeding 3000 fps using commercially available DMDs. The use of fiber-coupled RGB LDs as the light source allows the optical system to be compact and simplifies the alignment of its components. The generated holograms are converted to Fourier transform holograms, and a short-range diffraction calculation is applied to separate the imaging plane from the focal point of the zeroth-order light on the optical axis. The wavelength dispersion of the DMD is compensated for by superimposing plane waves on the hologram, which means that no special optical system is required. Moreover, the superimposition of the plane waves allows the object, conjugate, and zeroth-order light to be separated according to the shift theorem of the Fourier transform, thus eliminating the need for an extensive 4f system.

The principles on which the proposed technique is based are explained in detail in the next section. In section \ref{sec:experiment}, the details of the experimental system are presented along with the results. The resources used in this research, such as the programs used for hologram generation and the LD driver schematics, are available on GitHub\cite{GitHub}.

\section{Methods}
\begin{figure}[ht!]
    \centering\includegraphics[width=0.8\columnwidth]{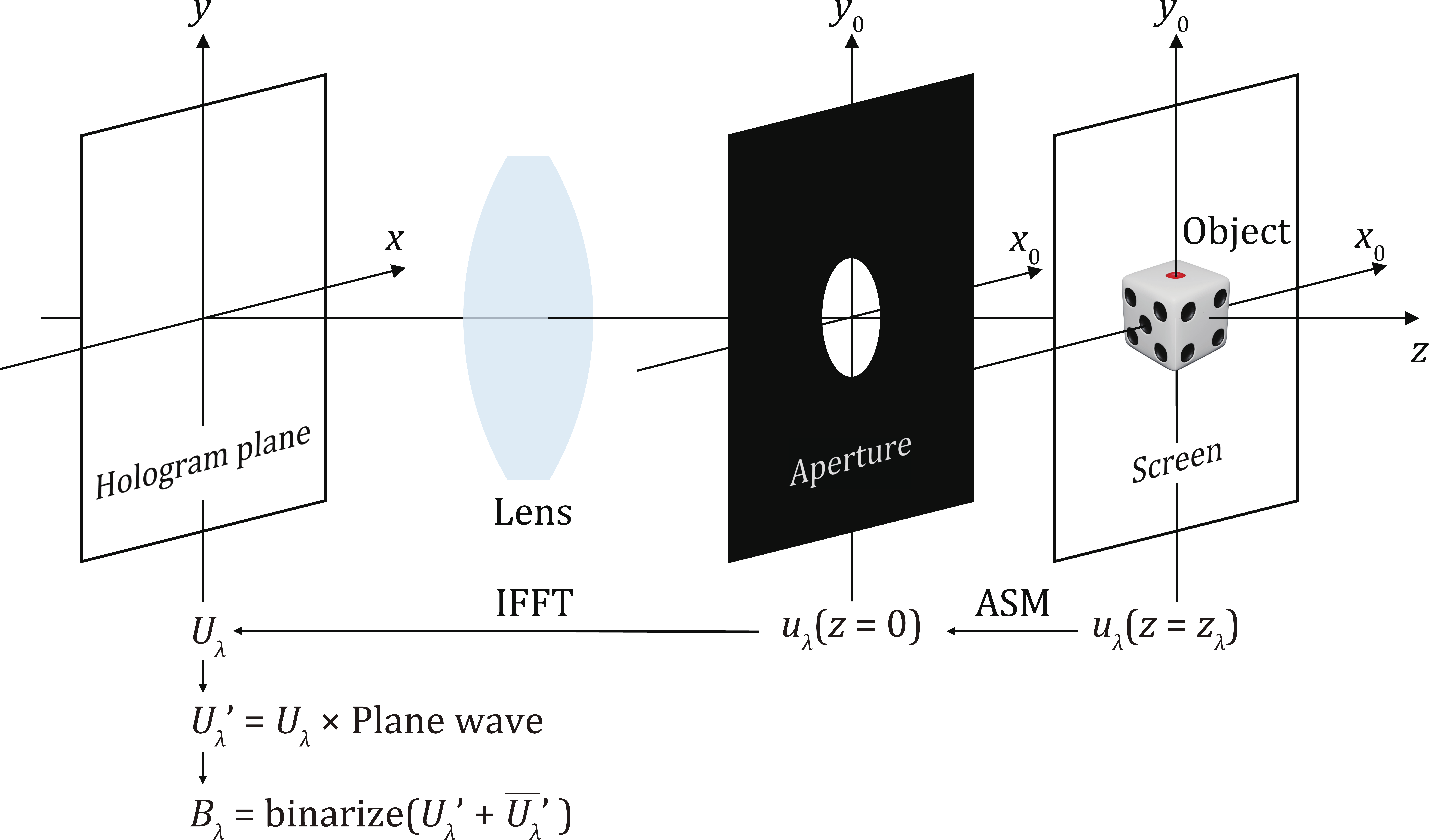}
    \caption{Overview of the optical system and hologram generation.}
    \label{fig:overview}
\end{figure}

This section describes the hologram generation method that was used in this study. An overview of the optical system and the hologram generation are shown in Fig. \ref{fig:overview}.

\subsection{Hologram generation}
\label{sec:generation}
The steps of the hologram generation are as follows.
\begin{enumerate}
    \item The image holograms at RGB wavelengths are generated using the polygon method.
    \item Short-distance diffraction calculations are applied to the image holograms.
    \item The image holograms are transformed into a Fourier transform hologram using an inverse fast Fourier transform (IFFT).
    \item A plane wave is superimposed to compensate for the wavelength dispersion.
    \item The obtained complex amplitude and its complex conjugate are added together to produce real numbers that are then binarized.
\end{enumerate}
First, the information corresponding to each color is extracted from the object model, and an image hologram, $u_{\lambda}$, is generated by the polygon method\cite{Matsushima2005,Matsushima2009,Pan2013,Zhang2022}. Since the size of the reconstructed image varies with the wavelength, it is necessary to scale the object model based on the appropriate lateral magnification:
\begin{equation}
    u_{\lambda}\pqty{x_{0},y_{0};z=z_{\lambda}} = \operatorname{polygon}\bqty{\operatorname{scale}\pqty{\mathrm{Obj},\beta_{\lambda}},\lambda}.
\end{equation}
Here, $\operatorname{scale}$ is a function used to scale the object model, $\operatorname{polygon}$ is a function used to generate the image hologram, $\mathrm{Obj}$ is the object model, and $\beta_{\lambda}$ is the lateral magnification at the wavelength $\lambda$. Since the maximum size of the reconstructed image is $\lambda l/p$, the maximum value of $\beta_{\lambda}$ is
\begin{equation}
    \beta_{\lambda} = \frac{\lambda l}{pL},
\end{equation}
where $p$ is the pixel pitch of the SLM, $L$ is the size of the SLM display area, and $l$ is a distance that depends on the type of optical system. The values of $l$ for each of the optical systems represented in Fig. \ref{fig:optics} are shown in Table \ref{tab:lengths}. In the case of the optical systems shown in Fig. \ref{fig:optics} (b) and (c), the lateral magnification can be adjusted by varying $l$.
\begin{table}[ht!]
    \centering
    \caption{Values of $l$ for different Fourier transform optics. $f$ is the focal length of the lens\cite{Goodman2017a}. The relation $1/z_{1} + 1/z_{2} = 1/f$ holds for the optical system shown in Fig. \ref{fig:optics} (c).}
    \begin{tabular}{cc}
        \hline
        Optical system & $l$ \\
        \hline
        Fig. \ref{fig:optics} (a) & $f$ \\
        Fig. \ref{fig:optics} (b) & $r$ \\
        Fig. \ref{fig:optics} (c) & $z_{2}\pqty{z_{1}-r}/z_{1}$ \\
        \hline
    \end{tabular}
    \label{tab:lengths}
\end{table}
When the in-plane size of the image reconstructed by the optics is $D$, Obj has to be scaled to $D/\beta_{\lambda}$ using the scale function. In the next step, a short-distance diffraction calculation is applied to $u_{\lambda}$ using the angular spectrum method (ASM)\cite{Goodman2017b} to separate the imaging plane from the focal point of the zeroth-order light on the optical axis:
\begin{equation}
    u_{\lambda}\pqty{x_{0},y_{0};z=0} = \operatorname{ASM}\bqty{u_{\lambda}\pqty{x_{0},y_{0};z=z_{\lambda}},-z_{\lambda}}.
\end{equation}
Here, $z_{\lambda}$ is the distance from the focal plane to the imaging plane. $\operatorname{ASM}$ is a function used to calculate the diffraction and is expressed in terms of $f_{x}$ and $f_{y}$, which are the spatial frequencies in the directions of the $x$- and $y$-axes, respectively:
\begin{equation}
    \operatorname{ASM}(u,z) = \operatorname{IFFT}\bqty{\operatorname{FFT}(u)H\pqty{f_{x},f_{y};z}}
\end{equation}
and
\begin{equation}
    H\pqty{f_{x},f_{y};z} = \exp\bqty{\mathrm{i}\frac{2\pi}{\lambda}\sqrt{1-\pqty{\lambda f_{x}}^{2}-\pqty{\lambda f_{y}}^{2}}z}.
\end{equation}
Since the longitudinal magnification of the optics for the wavelength $\lambda$ is $\alpha_{\lambda} = \beta_{\lambda}^{2}$, $z_{\lambda} = z_{0}/\beta_{\lambda}^{2}$ if the actual distance from the focal plane to the imaging plane is $z_{0}$. An IFFT is applied to the obtained wavefront:
\begin{equation}
    U_{\lambda}(x,y) = \operatorname{IFFT}\bqty{u_{\lambda}\pqty{x_{0},y_{0};z=0}}
\end{equation}
and plane waves are superimposed on the wavefront to compensate for the wavelength dispersion:
\begin{equation}
    U'_{\lambda}(x,y) = U_{\lambda}(x,y)\exp\bqty{2\pi\mathrm{i}\pqty{\frac{\alpha}{\lambda}x+\frac{\beta}{\lambda}y}},
\end{equation}
where $\alpha$ and $\beta$ are the direction cosines representing the direction of travel of the plane waves. The method for obtaining $\alpha$ and $\beta$ is described in section 2.3. The binary hologram $B_{\lambda}$ that is displayed on the DMD is obtained by adding the complex amplitude $U'_{\lambda}$ to its complex conjugate $\overline{U'_{\lambda}}$ to give a real number and then binarizing this number:
\begin{equation}
    B_{\lambda}(x,y) = \operatorname{binarize}\bqty{U'_{\lambda}(x,y) + \overline{U'_{\lambda}}(x,y)}.
\end{equation}
A set of full-color binary holograms is obtained by performing the above steps for each wavelength.

\subsection{Hologram reconstruction}
\begin{figure}[ht!]
    \centering\includegraphics[width=\columnwidth]{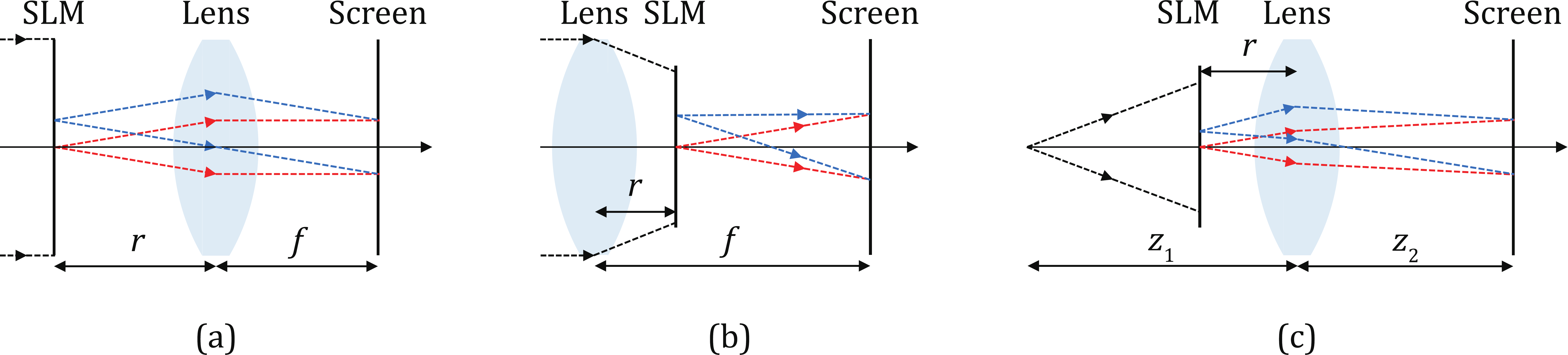}
    \caption{Fourier transform optics. The red and blue dashed lines represent diffracted light from the SLM. (a) The lens is placed in front of the SLM, and a plane wave is incident on the SLM. (b) The SLM is placed in front of the lens, and a plane wave passes through the lens before reaching the SLM. (c) The lens is placed in front of the SLM, and a spherical wave is incident on the SLM.}
    \label{fig:optics}
\end{figure}

This section explains how a hologram obtained using the method described in section \ref{sec:generation} can be reconstructed by the Fourier transform optical system shown in Fig. \ref{fig:optics}.
The complex amplitude $V_{\lambda}$ generated by the hologram $B_{\lambda}$ is
\begin{equation}
    V_{\lambda}(x,y) = U_{\lambda}(x,y)\exp\bqty{2\pi\mathrm{i}\pqty{\frac{\alpha}{\lambda}x+\frac{\beta}{\lambda}y}} + \overline{U_{\lambda}}(x,y)\exp\bqty{2\pi\mathrm{i}\pqty{-\frac{\alpha}{\lambda}x-\frac{\beta}{\lambda}y}} + U_{0}.
    \label{eq:amplitude}
\end{equation}
The terms on the right-hand side of Eq. (\ref{eq:amplitude}) represent the object, conjugate, and zeroth-order light, respectively; the zeroth-order light is produced by quantization errors in the binarization of the hologram and the unmodulated region of the DMD. The wavefront, $v_{\lambda}$, in the focal plane of the lens is
\begin{equation}
    \begin{split}
        v_{\lambda}\pqty{x_{1},y_{1}} &= A\pqty{x_{1},y_{1}}\iint_{-\infty}^{\infty}V_{\lambda}(x,y)\exp\bqty{-2\pi\mathrm{i}\pqty{\frac{x_{1}}{\lambda l}x+\frac{y_{1}}{\lambda l}y}}\dd{x}\dd{y}\\
        &= A\pqty{x_{1},y_{1}}\bqty{u_{\lambda}\pqty{\frac{x_{1}}{\lambda l}-\frac{\alpha}{\lambda},\frac{y_{1}}{\lambda l}-\frac{\beta}{\lambda};z=0} + \overline{u_{\lambda}}\pqty{-\frac{x_{1}}{\lambda l}+\frac{\alpha}{\lambda},-\frac{y_{1}}{\lambda l}+\frac{\beta}{\lambda};z=0} + \delta\pqty{x_{1},y_{1}}},
    \end{split}
    \label{eq:reconstruction}
\end{equation}
where $\delta$ is the delta function and $A$ is a second-order phase term that does not affect the intensity of the reconstructed image. Since $U_{0}$ is usually nearly uniform in space, its Fourier transform can be approximated by a delta function. The terms on the right-hand side of Eq. (\ref{eq:reconstruction}) correspond to the object, conjugate, and zeroth-order light, respectively, and are spatially separated in the $x_{1}$-$y_{1}$ plane. Since the object and conjugate light are distributed around $(l\alpha,l\beta)^{\intercal}$ and $(-l\alpha,-l\beta)^{\intercal}$ in the $x_{1}$-$y_{1}$ plane, respectively, a reconstructed image $u_{\lambda}(z = z_{\lambda})$ can be obtained by extracting only the object light using an aperture and propagating it by a distance $z_{0}$. The conjugate image $\overline{u_{\lambda}}(z = z_{\lambda})$ is formed in the plane $z = -z_{\lambda}$ as described by the relation $\operatorname{ASM}(\overline{u},-z)=\overline{\operatorname{ASM}}(u,z)$.

\subsection{Compensation for wavelength dispersion}
\begin{figure}[ht!]
    \centering\includegraphics[width=0.8\columnwidth]{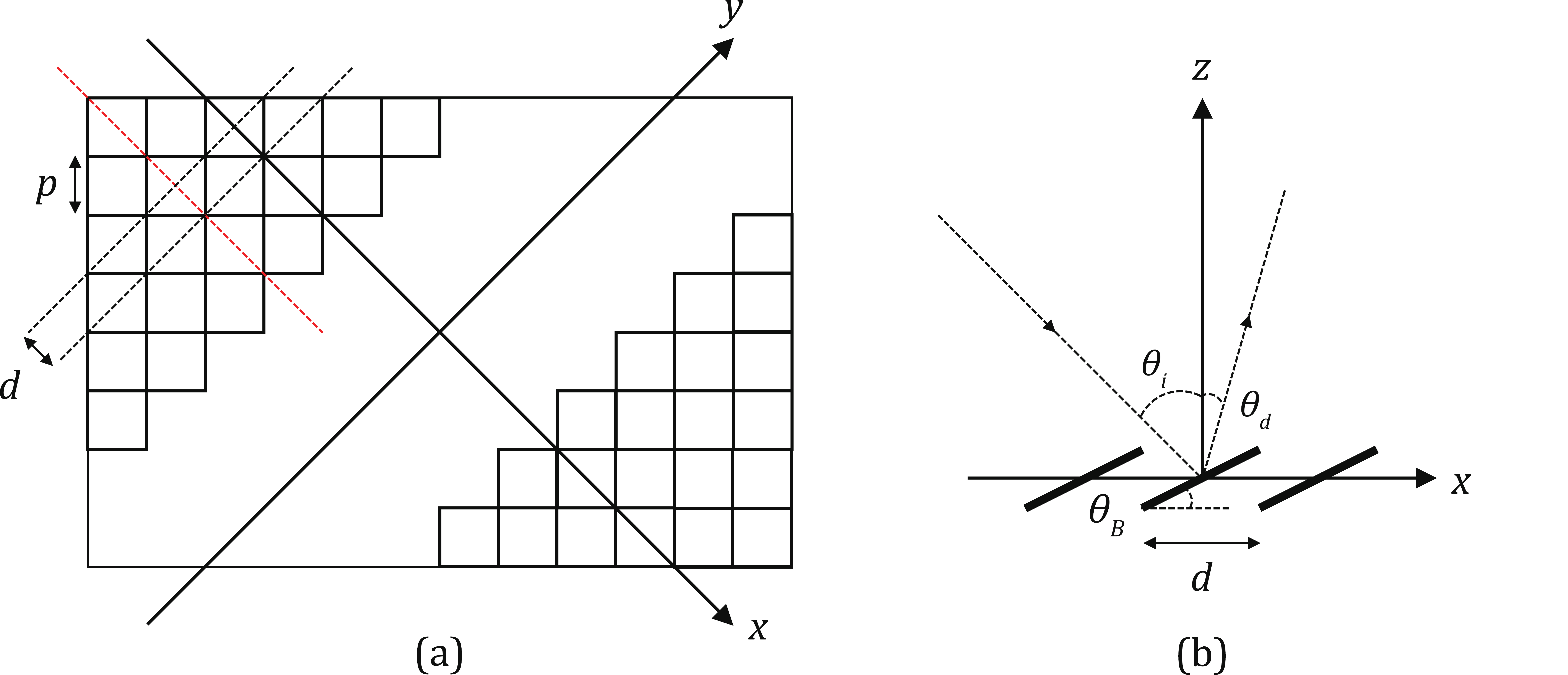}
    \caption{Structure of the DMD. Each micromirror tilts about the diagonal indicated by the dashed black line. When viewed along the dashed red line, the DMD has a blazed grating structure, as shown in (b). The pixel pitch of the DMD used in this study was $p = \SI{7.56}{\micro\meter}$, and the grating period along the diagonal line was $d = p/\sqrt{2} = \SI{5.35}{\micro\meter}$.}
    \label{fig:DMD}
\end{figure}

As shown in Fig. \ref{fig:DMD}, because of the DMD's structure, the DMD functions as a blazed grating, and the angle through which light is diffracted by the DMD varies with the wavelength. Therefore, it is necessary to compensate for wavelength dispersion to reconstruct RGB holograms at the same angle. There are two main approaches to compensating for the wavelength dispersion: one is to adjust the incident angle for each wavelength\cite{St-Hilaire1992,Lin2021,Chlipala2019,Takaki2015,Matsumoto2017}, and the other is to compensate with the grating phase\cite{Li2022}. In this study, the latter approach was used and plane waves were superimposed on the holograms.
If the lattice period is $d$ and $m$ and $n$ are the diffraction orders along the directions of the $x$- and $y$-axes, respectively, then the diffraction condition for a blazed diffraction grating can be expressed as\cite{Palmer2020}
\begin{equation}
    \vb{k}_{i} + \vb{k}_{d}(m,n) = \frac{2\pi m}{d}\vb{e}_{x} + \frac{2\pi n}{d}\vb{e}_{y},
    \label{eq:grating}
\end{equation}
where $\vb{k}_{i}$ and $\vb{k}_{d}$ are the wavenumber vectors of the incident and diffracted light, respectively, and $\vb{e}_{x}$, and $\vb{e}_{y}$ are the unit vectors along the $x$- and $y$-axes, respectively. Let the $z$-axis be the direction of the normal to the DMD display surface, the $y$-axis be the tilt axis, the $x$-axis be the axis perpendicular to the $y$-axis, and the $z$-$x$ plane be the incidence plane. If the incident and diffraction angles are $\theta_{i}$ and $\theta_{d}$, respectively, Eq. (\ref{eq:grating}) can be written
\begin{equation}
    \sin\theta_{i} + \sin\theta_{d} = \frac{m\lambda}{d}.
    \label{eq:diffraction}
\end{equation}
The reflected light intensity is a maximum in the direction of specular reflection from the micromirror, and the wavenumber vector, $\vb{k}_{r}$, of the reflected light is then
\begin{equation}
    \vb{k}_{r} = \vb{k}_{i} - 2\pqty{\vb{k}_{i}\vdot\vb{n}}\vb{n},
\end{equation}
where $\vb{n}$ is the vector corresponding to the normal to the micromirror. Assuming the incidence plane to be the $z$-$x$ plane, as in Eq. (\ref{eq:diffraction}), we have the condition
\begin{equation}
    \theta_{i} - \theta_{r} = 2\theta_{B},
    \label{eq:reflection}
\end{equation}
where $\theta_{r}$ and $\theta_{B}$ are the reflection and blaze angles, respectively. The blaze angle of the DMD used in this study was $\theta_{B} = \ang{12}$. When the direction of the hologram reconstruction is normal to the DMD, the condition for the maximum intensity of the diffracted light is $\theta_{d} = \theta_{r} = \ang{0}$, so the corresponding incident angle would then be $\theta_{i} = 2\theta_{B} = \ang{24}$. The actual diffraction angle, $\theta_{d}$, takes slightly different values depending on the wavelength, as shown below. The order, $m_{\mathrm{max}}$, at the maximum diffracted light intensity can be obtained from Eqs. (\ref{eq:diffraction}) and (\ref{eq:reflection}) as
\begin{equation}
    m_{\mathrm{max}} = \operatorname{round}\pqty{\frac{d}{\lambda}\sin2\theta_{B}},
\end{equation}
where $\operatorname{round}$ is the rounding function. The diffraction angle for $\theta_{i} = 2\theta_{B}$ is
\begin{equation}
    \theta_{d} = \sin^{-1}\pqty{\frac{m_{\mathrm{max}}\lambda}{d} - \sin2\theta_{B}}.
\end{equation}
For the RGB LDs used in this study, which had wavelengths $\lambda = 448, 518$, and \SI{638}{\nano\meter} and a grating period $d = \SI{5.35}{\micro\meter}$, the values of $m_{\mathrm{max}}$ and $\theta_{d}$ are as shown in Table \ref{tab:values}.
\begin{table}[ht!]
    \centering
    \caption{Values of $m_{\mathrm{max}}$ and $\theta_{d}$ for the wavelengths of the LDs used in this study.}
    \begin{tabular}{rrr}
        \hline
        $\lambda$ (\si{\nano\meter}) & $m_{\mathrm{max}}$ & $\theta_{d}$ (\textdegree) \\
        \hline
        $448$ & $5$ & $0.68$ \\
        $518$ & $4$ & $-1.11$ \\
        $638$ & $3$ & $-2.81$ \\
        \hline
    \end{tabular}
    \label{tab:values}
\end{table}
The optical axis of the incident light is along $2\theta_{B}$, and the diffracted light is directed along the angles shown in Table \ref{tab:values} for each wavelength. When the object light is reconstructed at the position $(X,Y)^{\intercal}$ in the $x_{1}$-$y_{1}$ plane, the corresponding direction cosine is $(\alpha,\beta,\gamma) = \pqty{X/l,Y/l,\sqrt{1-(X/l)^{2}-(Y/l)^{2}}}^{\intercal}$, as given by Eq. (\ref{eq:reconstruction}). However, the optical axis is tilted by $\theta_{d}$ from the $z$-axis around the $y$-axis. Hence, the direction cosine of the plane wave superimposed on $U_{\lambda}$ becomes $\vb{R}_{y}(-\theta_{d})(\alpha,\beta,\gamma)^{\intercal}$, where $\vb{R}_{y}$ is a matrix describing rotation around the $y$-axis. The direction cosine is then $\vb{R}_{z}(\ang{45})\vb{R}_{y}(-\theta_{d})(\alpha,\beta,\gamma)^{\intercal}$, with $\vb{R}_{z}$ being the matrix describing rotation around the $z$-axis when the DMD is installed at a \ang{45} tilt, as in the optical system shown in Fig. \ref{fig:setup}.

\section{Experiments}
\label{sec:experiment}
\subsection{Setup}
\begin{figure}[ht!]
    \centering\includegraphics[width=\columnwidth]{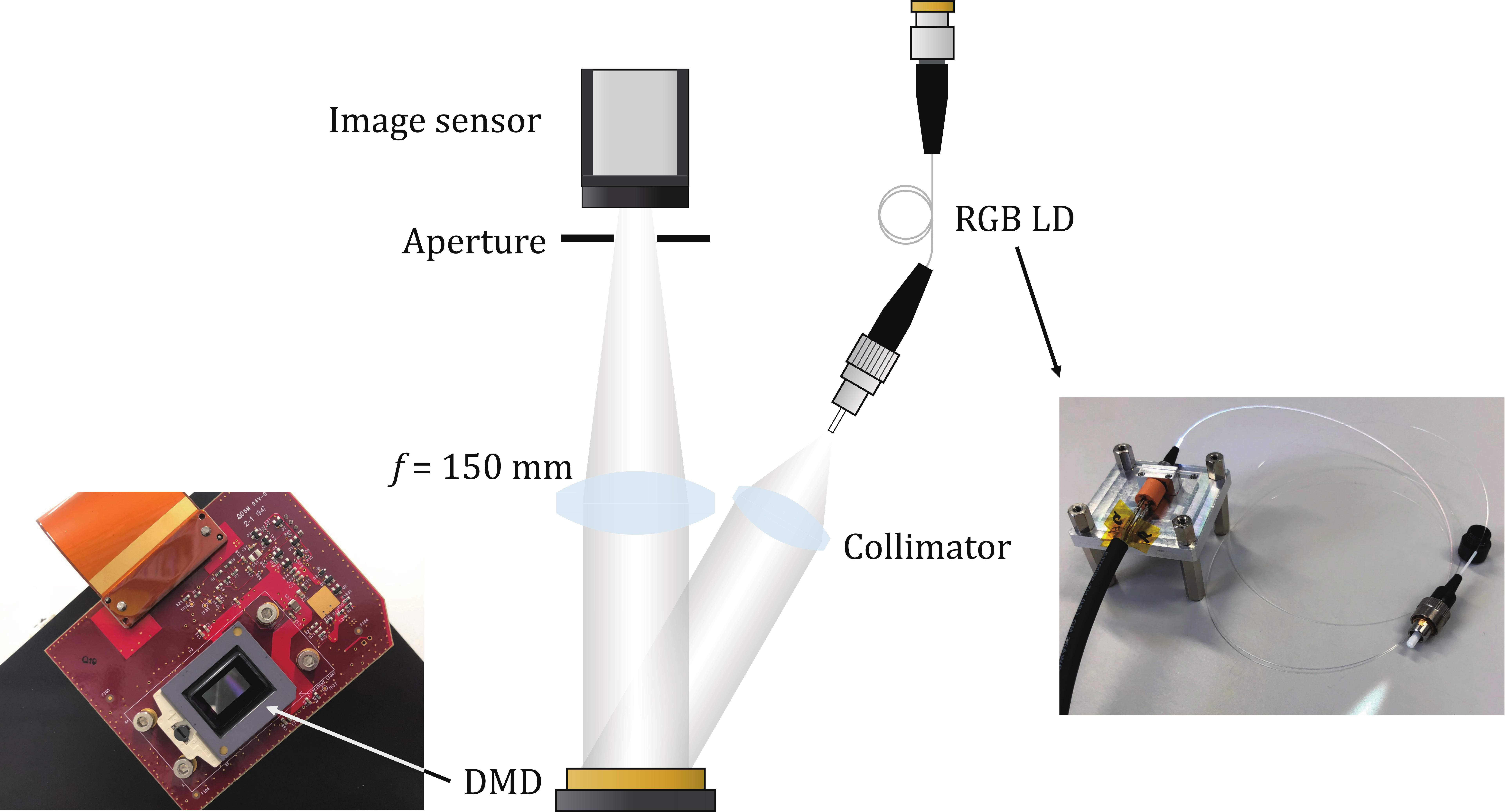}
    \caption{CGH image reconstruction optics, DMD, and light source. This arrangement corresponds to the Fourier transform optical system shown in Fig. \ref{fig:optics} (a). A Sumitomo Electric Industries SLM-RGB-T20-S1-2 was used as the fiber-coupled RGB LD. A simple schematic of the LD driver, which uses an operational amplifier and a MOSFET, is available on GitHub\cite{GitHub}.}
    \label{fig:setup}
\end{figure}

Figure \ref{fig:setup} shows the CGH image reconstruction optics and the fiber-coupled RGB LD light source used in this study.
In the setup shown above, the light emitted from the RGB LD was collimated by a collimating lens and entered the DMD at an angle of approximately $2\theta_{B}$. The light modulated by the DMD was Fourier transformed by a lens with a focal length of \SI{150}{mm}, and the zeroth-order, and conjugate light were removed by an aperture in the focal plane. The object light passing through the aperture propagated a short distance and formed an image on the sensor. In this study, a DLPLCR6500EVM evaluation board from Texas Instruments was used as the SLM. The DMD on the DLPLCR6500EVM had a pixel size of \SI{7.56}{\micro\meter}, a resolution of $1920\times1080$ pixels, and a frame rate of \SI{9523}{\hertz} for binary patterns. The evaluation board was rotated \ang{45} around the normal direction so that the optical axis of the diffracted light was parallel to the laboratory bench. The micromirror was tilted by $\pm\ang{12}$ around the diagonal axis (\ang{45}), resulting in an effective grating period of \SI{5.35}{\micro\meter} in the direction of the $x$-axis. A Baumer VCXU-50C camera with a resolution of $2448\times2048$ pixels, a pixel size of \SI{3.45}{\micro\meter}, and a maximum fps of 73 was used to acquire the reconstructed images.

\subsection{Results}
\begin{figure}[ht!]
    \centering\includegraphics[width=0.8\columnwidth]{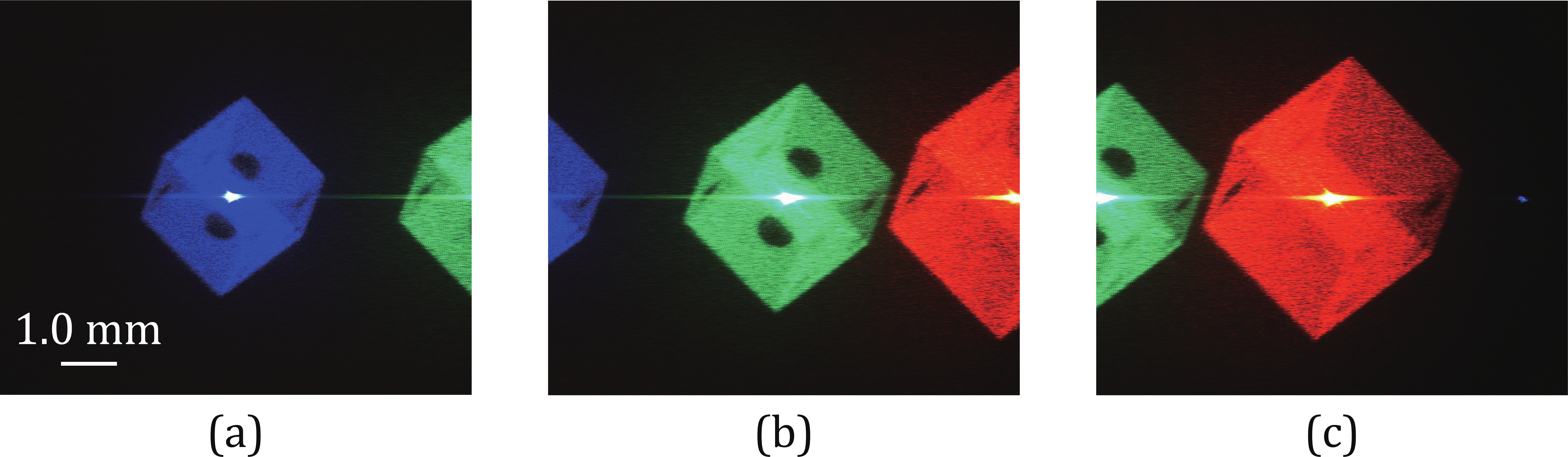}
    \caption{Reconstructed RGB holograms. (a), (b), and (c) are reconstructions with $\lambda = 448, 518$, and \SI{638}{\nano\meter}, respectively. The model scale and the DMD wavelength dispersion have not been compensated for.}
    \label{fig:dispersion}
\end{figure}

Fourier transform holograms were reconstructed without compensation for lateral magnification and wavelength dispersion to confirm the optical system's wavelength dependence. A reconstructed image of a dice is shown in Fig. \ref{fig:dispersion}. The distances between the R and G and G and B zeroth-order light are approximately 4.1 and \SI{4.8}{\milli\meter}, respectively, which is generally consistent with the diffraction angles shown in Table \ref{tab:values}. The diagonal lengths of the reconstructed RGB images are approximately 2.5, 2.0, and \SI{1.7}{\milli\meter}, respectively, and are proportional to the wavelength. The object, conjugate, and zeroth-order light have been superimposed on each other.

\begin{figure}[ht!]
    \centering\includegraphics[width=\columnwidth]{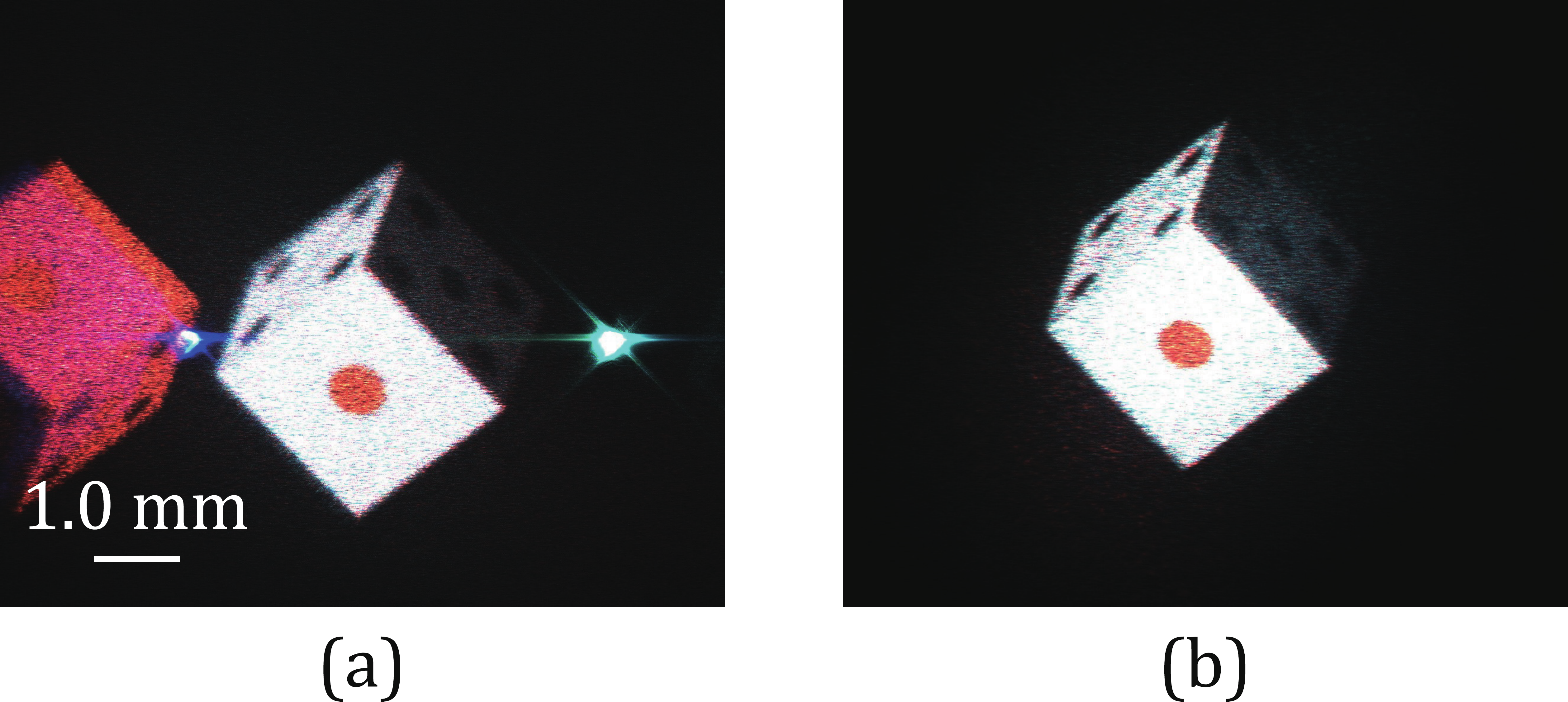}
    \caption{Reconstructed holograms compensated for lateral magnification and wavelength dispersion (a) without the application of the short-distance diffraction calculation and removal of the zeroth-order light and (b) with the application of the distance diffraction calculation and removal of the zeroth-order light.}
    \label{fig:color}
\end{figure}

Figure \ref{fig:color} shows reconstructed holographic images with compensation for lateral magnification and wavelength dispersion. In this case, the imaging plane was \SI{80}{\milli\meter} from the Fourier transform plane, the reconstruction direction was normal to the DMD, and the display interval for the holograms was \SI{105}{\micro\second}. A time of \SI{315}{\micro\second} was required to display a set of RGB holograms, resulting in an effective frame rate of approximately 3175 fps. The enable signal output from the evaluation board controlled the RGB LD switching. The color of the reconstructed image was adjusted by the drive current of the RGB LDs. The results show that the red, green, and blue colors can be appropriately synthesized by scaling the lateral magnification and by compensating for the wavelength dispersion. Speckle noise can be observed in the reproduced images. Since the hologram has a high display rate, the proposed method can maintain a sufficient frame rate even when the speckle noise is suppressed by application of the time-averaging method.

\begin{figure}[ht!]
    \centering\includegraphics[width=\columnwidth]{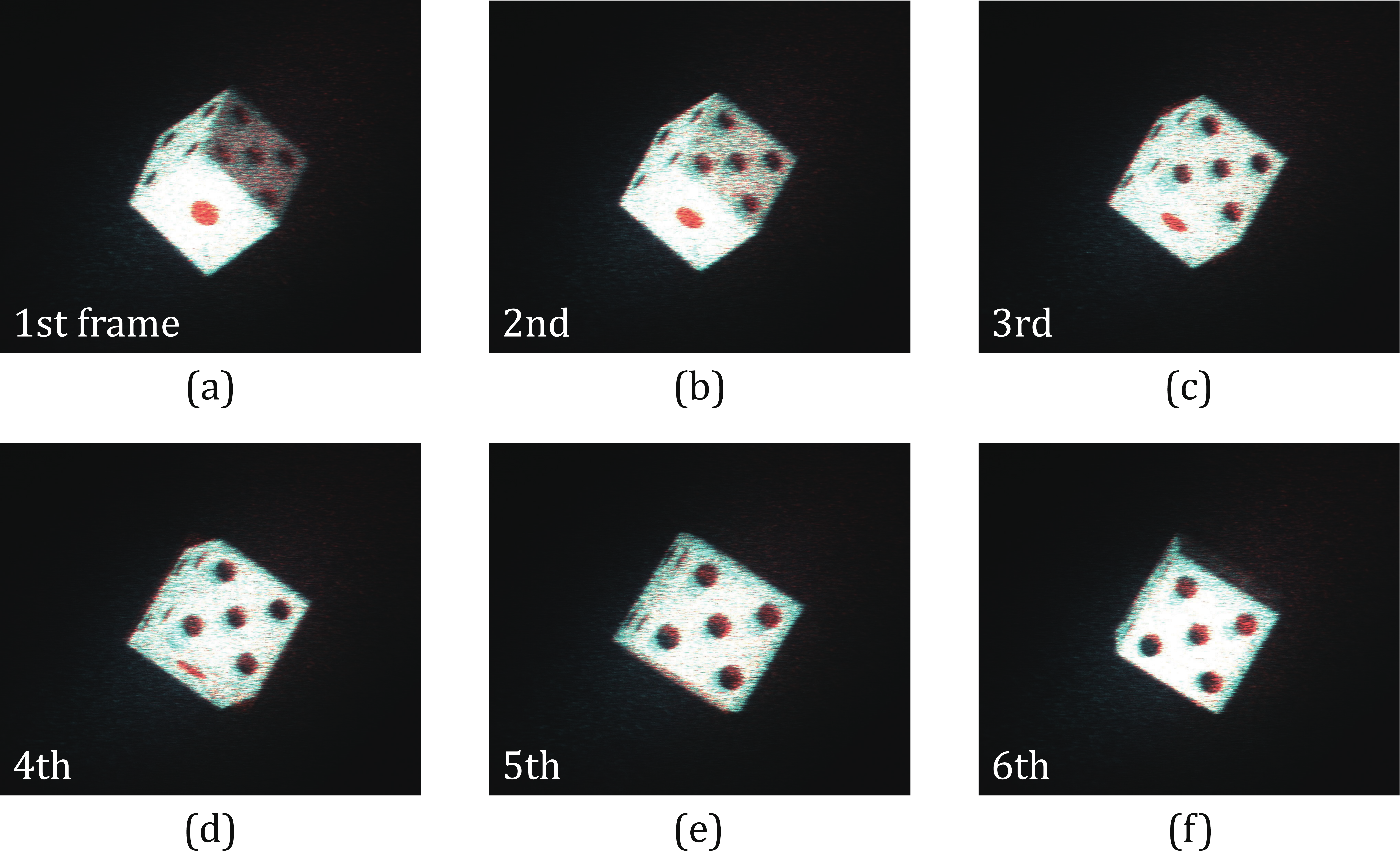}
    \caption{Reconstructed first six frames of a CGH video produced using the proposed method (see Visualization 1).}
    \label{fig:video}
\end{figure}

Figure \ref{fig:video} shows frames from a reconstructed CGH video produced by displaying holograms at high speed. The RGB holograms were produced by rotating the dice model by \ang{3} between each frame. The display interval was set to \SI{4630}{\micro\second} and synchronized with the image sensor frame rate of 72 fps. The dice appear to rotate smoothly, indicating that the proposed method can be used to effectively reconstruct full-color CGH videos.

\section{Conclusion}
In this paper, a method for producing high-speed full-color CGH displays using a DMD was proposed. The proposed method is based on a time-division multiplexing method. This enables full-color reconstruction with simple optics and removes unnecessary light by superimposing plane waves on the hologram and calculating the diffraction over a short distance, thus eliminating the need for a 4f system. Experiments showed that 3D video reconstruction is possible at over 3000 fps by taking advantage of the DMD's high-speed switching. This is more than 50 times faster than the widely used LC SLMs. The resources related to this research are available on the internet. The proposed method is expected to help improve the realism of CGH.

\section*{Acknowledgments}
This study used Wave Field Library (WFL) and Polygon Source Library (PSL) for computational hologram generation.


\bibliography{references}

\end{document}